# Kinetic Monte Carlo study of the type1/type 2 choice in apoptosis elucidates selective killing of cancer cells under death ligand induction

Subhadip Raychaudhuri

Indraprastha Institute of Information Technology, Delhi, Delhi 110020, India.

Email: subhadip@iiitd.ac.in

**Abstract**

Death ligand mediated apoptotic activation is a mode of programmed cell death that is widely used in cellular and physiological situations. Interest in studying death ligand induced apoptosis has increased due to the promising role of recombinant soluble forms of death ligands (mainly recombinant TRAIL) in anti-cancer therapy. A clear elucidation of how death ligands activate the type 1 and type 2 apoptotic pathways in healthy and cancer cells may help develop better chemotherapeutic strategies. In this work, we use kinetic Monte Carlo simulations to address the problem of type 1/ type 2 choice in death ligand mediated apoptosis of cancer cells. Our study provides insights into the activation of membrane proximal death module that results from complex interplay between death and decoy receptors. Relative abundance of death and decoy receptors was shown to be a key parameter for activation of the initiator caspases in the membrane module. Increased concentration of death ligands frequently increased the type 1 activation fraction in cancer cells, and, in certain cases changes the signaling phenotype from type 2 to type 1. Results of this study also indicate that inherent differences between cancer and healthy cells, such as in the membrane module, may allow robust activation of cancer cell apoptosis by death ligand induction. At the same time, large cell-to-cell variability through the type 2 pathway was shown to provide protection for healthy cells. Such elucidation of selective activation of apoptosis in cancer cells addresses a key question in cancer biology and cancer therapy.

**Introduction**

Death ligand induced apoptosis is a frequently used mode of cell death activation and is also important for its potential application in cancer therapy. Tumor necrosis factor related apoptosis inducing ligand (TRAIL), or recombinant forms of that ligand, has particularly attracted recent attention due to its role as a promising anti-cancer agent [1-4]. However, a clear mechanistic understanding of such selectivity in TRAIL induced apoptosis in cancer cells is still lacking. Expression levels of various pro- and anti-apoptotic molecules in the membrane proximal and other signaling modules (in the apoptotic pathway) may modulate selectivity in cancer cell apoptosis. In a more general manner, the systems level regulatory mechanisms (such as the type 1/type 2 choice) of death ligand induced activation of the apoptotic death pathway needs to be clearly elucidated. Presence of decoy receptors that are also capable of engaging the death ligands makes it difficult to understand the mechanisms of death ligand induced activation of the membrane proximal module. Moreover, two distinct clustering mechanisms have been indicated for TRAIL decoy receptors [2, 5] providing additional complexities into the problem.

Death receptor activation, typically by death ligand engagement, is the mode of apoptotic activation in a wide variety of cellular and physiological situations. Death ligand induced apoptosis has been implicated in the death of strongly self-reactive immature lymphocytes (also for peripheral tolerance) [6]. In addition, death ligand mediated apoptosis might be involved in maintaining homeostasis of mature lymphocytes that have undergone clonal expansion (upon antigenic stimulation) [6]. Killing of pathogen infected cells or tumor cells by cytotoxic T lymphocytes (CTLs) and natural killer (NK) cells can be mediated by death ligands [7]. TRAIL was found to be constitutively expressed on immature NK cells and also on NK cells of adult liver, and, was shown to be involved in NK cell mediated suppression of tumor metastasis [8]. NK cell's TRAIL / Fas mediated killing of tumor cells might play an important role in the elimination of cancer stem cells as depletion of NK cells led to a significant increase in cancer stem cell population [9]. Role of TRAIL is also getting increasingly recognized in cancer immunotherapy [10]. In the context of Alzheimer's disease (a common form of neurodegenerative disorder), amyloid beta (Aβ) oligomers and aggregates were shown to engage both the type 1 (extrinsic) and the type 2 (intrinsic) pathways of apoptosis [11-13]. Therefore, a clear elucidation of the death ligand induced apoptosis may have implications for various important biological problems.

In our previous studies, we addressed the problem of understanding systems level regulation of type 1/type 2 choice in apoptotic activation [14, 15]. Results obtained from our computational studies indicated that type 1/type 2 choice is regulated at a systems level, both in healthy and cancer cells. In addition, analysis of our *in silico* data provided insights [14] into recent experimental studies that explored the type 1/type 2 choice in apoptotic activation. In a recent study, we showed that the concentration of death ligand is a key determinant of type 1/ type 2 choice in healthy cells [15]. Whether death ligand concentration also impacts the type 1/type 2 choice in cancer cells (in a manner similar to healthy cells) remains to be investigated. Such elucidations seem to be crucial for designing cancer therapies that rely on activating the apoptotic pathways by death ligand (such as TRAIL) induction.

Experimental studies including genomic analysis of cancer cells have indicated that the genome and proteome of cancer cells are markedly different from that in healthy cells. Anti-apoptotic proteins are frequently over-expressed in cancer cells making them particularly resistant to apoptotic activation. How it is then possible to activate the apoptotic pathways selectively in cancer cells in such a manner that healthy cells remain mostly protected, remains a challenging question that needs to be addressed before successful anti-cancer therapies can be developed based on apoptotic induction. In our earlier studies we addressed the mechanisms for selective killing of cancer cells such as by inhibition of Bcl2 family anti-apoptotic proteins [16-19]. It seems likely that some of the pro-apoptotic molecules (such as Bid and Bax) are also over-expressed in certain cancer cells making them susceptible to apoptotic induction under Bcl2 inhibition [19]. However, over-expressions of anti- and pro-apoptotic molecules are cancer type specific (and possibly patient-specific), and, also varies within a tumor population. Thus studies need to be carried out to find suitable strategies for apoptosis induction in a specific case. Cancer cells in which p53 mediated pathway is defective become resistant to genotoxicity based chemotherapeutic treatments [20] but death ligand mediated apoptotic activation (possibly in combination with genotoxic or other types of chemotherapeutic treatments) still remains a viable option [21]. In many cancer types, TRAIL induction seems to induce apoptotic activation in cancer cells but normal cells are not significantly impacted [2]. In some of the earlier studies, TRAIL was found to be selective for cancer cells even though the receptor for TRAIL was present in many types of healthy cells [22, 23]. Differences in the expression levels of death and/or decoy receptors between cancer and healthy cells have been implicated in selective targeting of cancer cells [22-24]. Naturally found substances such as the medicinal fruit Phyllanthus emblica Linn. also seem to selectively activate the type 1 pathway in certain cancer cells [25]. Elucidating the precise mechanisms of such selective targeting of cancer cells through the death receptor mediated pathway is lacking but such elucidations may allow more effective cancer therapies. One needs to probe the mechanisms for: (i) death and decoy receptor engagement to death ligands that result in receptor clustering, (ii) selective activation of the membrane module in cancer cells, and, (iii) synergy between apoptotic vulnerability in the membrane module and that in type 1/type 2 pathways (in cancer cells).

In this work, we use kinetic Monte Carlo simulations to elucidate the single cell mechanisms of the type 1/type 2 choice in cancer cells *in silico*. Results obtained from this study indicates that over-expression of certain anti-apoptotic proteins may make cancer cells prone to either type 1 or type 2 activation. However, in most cases increased death ligand induction leads to an increase in type 1 activation, as also observed in our studies for healthy cells [15]. In some situations, increased death ligand concentration may actually change the signaling phenotype from type 2 to type 1. Consistent with earlier experimental studies, the DR / DCR ratio in the membrane module is shown to be a key determinant of the type 1/type 2 choice. Elucidations of type 1/type 2 choice in cancer cells also seem to provide crucial insights into the mechanisms of selective killing of cancer cells under TRAIL induction. In this context, single cell analysis of this study elucidates the role of cell-to-cell variability and slow activation in protecting healthy cells. I*n silico* approaches can be utilized to find optimal strategies for selective apoptotic activation in cancer cells under death ligand induction.

**Methods**

We used Monte Carlo simulations to carry out computational study of death ligand induced apoptosis in cancer cells. A simplified network model of apoptotic cell death signaling, which incorporates key regulators in both type 1 and type 2 pathways (Figure 1), was simulated *in silico*. The extent of type 1/type 2 activation was assessed in a quantitative manner at the level of single cells.

The signaling model for apoptotic cell death

In our computational model, induction of death ligands (such as FasL, TRAIL) activates the initiator caspases (caspase 8) in the membrane proximal death module and thus acts as a trigger for apoptotic activation. Once caspase 8 is activated in the membrane module it initiates signaling through both type 1 and type 2 pathways. Death ligands bind with death receptors with high (~ nanomolar) affinity causing oligomerization of death receptors and recruitment of adaptor molecules through interaction between their death domains (DD) [2, 26]. Binding of adaptor proteins with ligand-bound death receptors (through death domains DD) has been estimated to be of moderate affinity (~ $10^6$ M$^{-1}$) [27]. This affinity might be higher as these adaptor molecules bind with pro-caspase 8 with high affinity ~ $10^8$ M$^{-1}$. The adaptor molecules may also bind with free receptors but the binding affinity should be small so that spontaneous apoptotic activation (in absence of death ligands) does not occur in healthy cells (this binding affinity is set to zero in our current simulations). The clustered environment of death receptors and adaptor molecules is known as death inducing signaling complex (DISC) as it could recruit, cluster, and activate caspase 8. Death effector domains (DED) of inactive pro-caspase 8 molecules bind to DED of adaptor proteins and then get cleaved to active caspase 8 (when two procaspase 8 molecules are proximal). We also consider decoy receptors (not considered in our earlier studies) that lack death signaling capability as they cannot recruit adaptor proteins. Decoy receptors for TRAIL are known to inhibit formation of DISC by two distinct mechanisms: (i) for decoy receptor 1 (DCR1, also known as TRAIL-R3): by pre-clustering (likely in lipid raft domains [2]) and then recruiting significant fraction of death ligands in those micro-clusters and (ii) for decoy receptor 2 (DCR2, also known as TRAIL-R4): by clustering with DR5 in DISC, upon death ligand engagement, but preventing DR4 recruitment in DISC [2, 28]. DR4 (TRAIL-R1) and DR5 (TRAIL-R2) are two well-studied receptors for TRAIL. In the first scenario, recruitment of death ligands in DCR1-raft regions supposedly leads to reduction in available death ligands for DR4/DR5 binding and thus inhibit DISC formation. In the second case, decoy receptor DCR2 clustering with DR5 seems to diminish clustering of caspase 8 within DISC as decoy receptors do not recruit adaptor proteins. Clustering of receptors was modeled through an effective free energy parameter E, which could capture the effective attraction (reduction in energy) when two receptors occupied neighboring nodes. For free (unbound) death receptors (and decoy receptors) the energy parameter was assumed to be small ~ -$K_B$ T to prevent spontaneous pre-clustering of receptor molecules. However, pre-clustering in lipid raft domains has been observed for DCR1 due to their GPI anchor [2, 28]. We lowered the value of the free energy parameter to E = -3$K_B$ T to capture such ligand independent clustering of receptors. Ligand binding was assumed to enhance the effective attraction between neighboring receptors resulting in a lowering of the free energy parameter from its basal value. For initially unclustered free receptors, ligand binding lowered the value of the free energy parameter to a value E = -3$K_B$ T. For pre-clustered receptors, ligand binding resulted in E = -4$K_B$ T (further lowering of free energy). The free energy parameter E, the number of receptors and ligands, and, the ligand binding affinity were found to be the key parameters for inducing receptor clustering and DISC formation.

Active caspase 8 molecules engage both type 1 and type 2 pathways. In the type 1 pathway, caspase 8 directly activates the effector caspases (caspase 3/7). In the type 2 pathway, a second initiator caspase (caspase 9) is required for effector caspase activation and signal amplification. Initial routing of the activation signal through type 1/ type 2 pathways is governed by relative affinity of caspase 8 for pro-caspase 3 (type 1 pathway) and Bid (type 2 pathway) [15]. Final determination of the type 1/type 2 choice depends on how fast the activation signal propagates through each of those two pathways to finally activate caspase 3 [15]. In the type 2 pathway, signal propagation is regulated by the pre- and post-mitochondrial signaling modules. In the pre-mitochondrial module, pro- and anti-apoptotic Bcl2 family proteins determine the extent of Bax activation and cytochrome c release from mitochondria. In the post-mitochondrial module, cytochrome c binds with Apaf to form the multi-molecular cyto c-Apaf-ATP complex apoptosome. In the clustered environment of apoptosome, initiator caspase 9 is activated (similar to activation of caspase 8 within DISC), which in turn, activates the effector caspases. Apoptotic inhibitor XIAP binds to pro-caspase 9, caspase 9 and caspase 3, thus inhibits effector caspase activation in both type 1 and type 2 pathways. Activation of effector caspases (represented by caspase 3 in our model) was taken as a downstream readout of apoptotic cell death signaling and the time-course of its activation was measured (at the level of single cells) in our MC simulations.

Earlier studies indicated that the type 1/type 2 choice is death receptor specific [15, 29] as one type of death ligand activates a specific death receptor (or a few specific receptors) and different death receptors may have different expression levels for a given cell type. It is reasonable to assume that decoy receptors are also ligand specific and their expression level varies depending on the decoy receptor considered. DCR1 and DCR2 (TRAIL-R3 and TRAIL-R4), for example, are specific for TRAIL and not relevant for FasL induced apoptosis [5]. Therefore, it should be noted that the death receptors/ligands simulated here capture the effect of only one type of receptor/ligand pairs (such as Fas/FasL or DR4-DR5/TRAIL), not the combined effect of all the death receptors/ligands on a given cell type. As we include DCR 1/DCR 2 in our current simulations results are most relevant for TRAIL induced apoptosis.

Even though we simulate some major components of the apoptotic death regulatory pathway that are essential for addressing the type 1/type 2 choice in cancer cells, it has several simplifying assumptions. In our model, functionally similar molecules (pro- or anti-apoptotic) are coarse-grained by a representative protein. Bcl-2 (B cell lymphoma protein 2), for example, represents all the Bcl-2 family proteins (such as Bcl-2, Bcl-xL, Mcl-1) with similar anti-apoptotic properties. The present model do not consider the following: (i) synthesis and degradation of signaling molecules [30] and (ii) engagement of other pathways (such as NFκβ) by components of the apoptotic death pathway [2, 31].

Hybrid Monte Carlo (MC) simulation of apoptotic cell death signaling

We developed a hybrid Monte Carlo simulation that combines the following two approaches: (1) a probabilistic rate constant based (implicit free energy) kinetic Monte Carlo simulation for various reaction moves such as diffusion, binding/unbinding and catalytic cleavage; (2) an explicit free-energy based kinetic Monte Carlo model that captures clustering of receptor molecules utilizing energy function based diffusion moves. At each Monte Carlo (MC) step, one molecule was sampled (on average) once to allow for either diffusion or a reaction move. Acceptance/Rejection of the moves was carried out based on either through the probability constants (in implicit energy MC) or by the Metropolis criteria (in explicit energy MC) and ensures detailed balance is satisfied.

A simulation volume of $1.2 \times 1.2 \times 1.2$ μm$^3$ (corresponding to a $60 \times 60 \times 60$ lattice with lattice spacing $\Delta x = 20$ nm) is chosen in such a manner that the number of molecules (for each molecular species) is equal to the nanomolar concentration. Death and decoy receptors are placed on one surface of the simulation lattice where they engage death ligands placed on a surface parallel to death / decoy receptor surface. Cytochrome c /Smac is initially localized in an $18 \times 18 \times 18$ sub-lattice within the intracellular volume. Each MC step ($\Delta T$) is chosen to be $10^{-4}$ s based on known mobility of cytosolic molecules [15]. A typical simulation is run for $10^8$ MC steps. Kinetic reaction rates (such as $k_{on}$ /$k_{off}$) and molecular concentrations were obtained from values reported in the literature (mostly from [27])[32] and utilized in our previous work [14, 15, 19, 33] (unless specified otherwise). In our studies of cancer cells, concentrations of over-expressed molecules are provided (fold over-expression was defined with respect to concentrations used in simulating apoptosis in healthy cells). In simulating selective apoptotic activation in cancer cells the following death receptor levels were used: 100 (for cells with type 1 susceptibility) and 20 (for cells with type 2 susceptibility). We used 20 molecules for the decoy receptors (not considered in our earlier studies) except for one set of simulations where it was set to zero. Death ligand concentration was varied in our simulations to study the effect of its variation on the type 1/type 2 choice. The number of death adaptor proteins and cFLIP was taken to be 30 (unless stated otherwise) [27, 32, 34], which is comparable to the expression level of pro-caspase 8 molecules used in our simulations. Such a kinetic MC scheme allows rigorous study of the type 1/type 2 choice problem, which is essentially a dynamical problem. Details of our simulation method can be found in our earlier works.

Single cell data analysis

Each run of our Monte Carlo simulation corresponds to apoptotic activation in a single cell, thus, data obtained from Monte Carlo runs can capture cell-to-cell stochastic variability (including inherent variability) in signaling activation. Each run of the simulation corresponds to activation at a single cell level. Statistical analysis was carried out using data obtained from many single cell runs (64 runs were used in current simulations). We define the following parameters for quantitatively assessing the extent of type 1 and type 2 activations: $N_{type1}/(N_{type1} + N_{type2})$ and $N_{type2}/(N_{type1} + N_{type2})$, where $N_{type1}$ and $N_{type2}$ represents the number of active caspase 3 molecules generated by caspase 8 and caspase 9, respectively. In our simulations, the type 1 (or type 2) activation was estimated at a time when caspase 3 molecules undergone half-maximal activation. Average and standard deviation (measure of fluctuation) in type 1/ type 2 activation fraction was estimated based on 64 single cell runs. At the level of single cells, a cell was called a type 1 cell if $N_{type1}/(N_{type1} + N_{type2}) > 0.5$ (more than 50% of activation was through the type 1 pathway).

**Results**

Increasing death ligand concentration increases type 1 activation in cancer cells

Concentrations of apoptotic and anti-apoptotic molecules in the death signaling pathway is cell type specific. In cancer cells, such cell type specific variability is presumably much more pronounced as cancer cells found to exhibit genomic instability and aberrant expression of both pro- and anti-apoptotic molecules (in a cancer-type specific manner). Such variability may also exist among genetically heterogenous sub-populations within a given tumor cell [35, 36]. However, some of the key anti-apoptotic molecules such as Bcl2 like proteins and XIAP are over-expressed in a large number of cancers [18, 37] and incorporated in our simulations of typical cancer cells. We also assume moderate over-expression of apoptotic Bid and Bax molecules (in many cancers inhibition of Bcl2 like proteins results in mitochondrial cytochrome c release and apoptosis activation) [19]. We study the effect of variation in death ligands on apoptotic activation (type 1/type 2 choice) in such representative cancer cells.

When DL concentration is low (~ 2 molecules on $1.2 \times 1.2$ $\mu m^2$ cell surface), very high XIAP expression (~ 3 fold or higher) suppresses type 2 apoptosis as XIAP strongly inhibits caspase 9 activation. Only weak type 1 activation was observed under such low DL concentrations. Increased DL concentration resulted in markedly increased apoptosis and also increased activation through the type 1 pathway (Figure 2). However, the type 1 activation seemed to proceed only after Smac was released from mitochondria and XIAP was inhibited [14, 38, 39]. In this type 2 assisted type 1 mode of activation, the time to cytochrome c/Smac release exhibited significant stochastic variability mainly due to cell-to-cell stochastic fluctuations in activation of the membrane module. In contrast, when cancer cells had regular XIAP levels but decreased death adaptor proteins and increased cFLIP, death ligands induced weak type 2 activation (Figure 3) with its characteristics large cell-to-cell variability (Figure 4). Combined effect of decoy receptors, lower adaptor protein levels and higher cFLIP was to impair DISC formation and pro-caspase 8 processing.

Such weak type 2 activation might explain resistant cancer cells such as DCR2 expressing cells of MCF-7 breast cancer cell line [31]. Here also, increased DL concentration resulted in increased fraction of type 1 activation (Figure 3). In corresponding healthy cells (when over-expressions of pro- and anti- apoptotic molecules are removed), low level of death ligand induction mainly activated the type 2 pathway but the activation was switched to type 1 dominant mode with increasing ligand concentrations (Figure 5) [15]. For healthy cells equipped with low level (~ 10 molecules) of adaptor proteins and comparable amount of death and decoy receptors (as in Figure 5), the decoy receptors DCR2 [31] had a stronger anti-apoptotic effect than DCR1 when intermediate level (~ 10 molecules) of death ligand was used in our simulations. In Figure 6, we show distinct clustering mechanisms for DCR1 and DCR2 and this may explain differences in apoptotic activation observed between DCR1 and DCR2. In cancer cells, when Bcl2 is 10 fold overexpressed along with moderate over-expression (~ 5 fold) of Bid (BH3 only proteins) and Bax / Bak, increased concentrations of death ligand induction change the activation from type 2 dominant mode to type 1 dominant mode (Figure 7).

Mechanisms for selective killing of cancer cells by TRAIL induction (apoptotic vulnerability in the membrane module of cancer cells)

Genome and proteome of cancer cells are markedly different from that in healthy cells and such differences may allow selective targeting and elimination of cancer cells. In many cases, the inherent state of the membrane module (including the death receptor expression) in cancer cells is more susceptible to apoptotic activation than that in healthy cells. Recent experiments indicated that expression of death receptors DR4 and/or DR5 (TRAIL receptors) can be high in certain cancer cells [31, 40]. In addition, DR4/DR5 and other death receptors can be upregulated by chemotherapeutic agents and also by genetic mechanisms [40-43]. In addition, death adaptor proteins might have lower expression levels in healthy cells (such as that are prone to type 2 activation [27]) and/or its expression can be induced selectively in certain cancer cells [40]. In a more general manner, it might be possible to alter the inherent proteomic and lipidomic state of the plasma membrane by chemotherapeutic treatments. For example, XIAP inhibitor embelin has been found to induce down-regulation of sFLIP in malignant glioma cells [41]. Based on this observation we simulated apoptosis activation *in silico* for cancer cells showing highly expressed DR4/DR5 (~ 100 molecules) [41-43] or Apaf proteins (~ 100 nanomolar) [44], as well as downregulated cFLIP (~ 10 nanomolar) [41]. In corresponding normal cells expression levels of DR4 /DR5 and Apaf were assumed to be low-to-moderate. Significant expression of cFLIP (~ 30 nanomolar) was used for healthy cells. As discussed in the methods section, two different scenarios for decoy receptors (for DCR1 and DCR2) were simulated for TRAIL induced clustering of DR4/DR5.

We considered two possible strategies for selectively activating the death receptor induced pathway in cancer cells. The type 2 pathway can be activated in cells for which DR expression is not very high (DR / DCR ~ 1), Bid and Bax type proteins are over-expressed to some extent, apoptotic inhibitors Bcl2 and XIAP are either not highly expressed or those can be inhibited (such as by embelin alone [45] or by combined treatment of ABT-737 and embelin [18]), and, Apaf expression is significant (compared with healthy cells) [44]. In contrast, when the type 2 pathway remains strongly resistant to activation (such as due to low Bid-Bax levels and high Bcl2 like inhibitors) but large concentration of death receptors is either present or can be induced, activating the type 1 pathway turns out to be a better strategy. In either case, we observe selective killing of cancer cells while activation in healthy cells remain low. We present quantitative estimations of type 1 and type 2 activations (based on population averaged fraction of type 1 / type 2 activation over 64 single cell runs; see Methods). We can also assign a type 1/type 2 activation phenotype to each cell that has undergone sufficient effector caspase activation. For a single cell, if the type 1 activation exceeds 50% of total, then the cell is called a type 1 cell (see Methods).

(i)  Selective killing of cancer cells under high affinity death ligand induction

In certain cancer cells the adaptor protein expression can be higher or it can be induced by genetic mechanisms (such as done for neural cancer cells [40]). In such a scenario, it seems to be possible to selectively activate the apoptotic pathway in cancer cells by application of a very low concentration of death ligands (~ 2 molecules/1.44 $\mu m^2$). Even when the adaptor protein level in cancer cells is comparable to that in healthy cells it might still be possible to activate apoptosis in cancer cells using low concentration of death ligands (such as when DR/DCR ratio is high). In our simulations, DL = 2, DR = 10, DCR = 20 and adaptor = 10 (molecules) resulted in very weak activation in healthy cells (except for activation in a small number of cells with majority of cells getting activated through the type 2 pathway) (Figure 8). For cancer cells, we assumed a 3-fold higher expression level for the death adaptor protein. As before, when the type 2 pathway is susceptible for activation in a cancer cell type (due to overexpressed Apaf and low/inhibited Bcl2), DR = 20, DCR = 20 and DL = 2 resulted in strong apoptotic induction (67% for DCR1 and 80% for DCR2 as shown in Figure 8) in the simulation time-scale of ~ 3 hrs. In contrast, when the DR4/DR5 concentration is high, moderately rapid activation of the type 1 pathway is possible even for low death ligand concentrations DL = 2 (Figure 8). The ratio DR/DCR emerges as a key parameter that determines the extent of activation in the membrane proximal module as well as the extent of cell-to-cell stochastic variability in caspase 8 activation. In certain cases the type 1/type 2 choice at the level of single cells may differ from that estimated based on average over many cells. In our simulations for cancer cells equipped with high level of Apaf (type 2 susceptibility), type 1/type 2 choice estimated at the level of single cells (Table 1) differs from the estimation based on average type 1/type 2 activation (Figure 8). Clearly, the type 1/type 2 choice in cancer cells is regulated at a systems level and is dependent on the inherent state of various signaling modules (Figure 1) in the apoptotic pathway.

We also considered the scenario where negligible amount of decoy receptors were present on cancer cells, as indicated by some earlier experimental studies [22, 23]. When DCR concentration was set to zero in our simulations, all of the single cells undergone nearly complete activation even for moderate level of death receptor (~ 20 molecules) and adaptor protein (~ 10 molecules) presence and low level (~ 2 molecules) of death ligand induction (data not shown). Such a result should also be relevant for FasL (CD95L) induced apoptosis if the selectivity is achieved for cancer cells or local application of FasL is possible. We also observed significant decrease in cell-to-cell variability in casapse 8 activation when DCR = 0 and a stochastic-to-deterministic transition might be possible to achieve through the type 2 pathway [19].

**Table 1.** Type 1/Type 2 choice in cancer cells (under DL = 2 induction) at the level of single cells (based on 64 single cell runs)

| Type of cell | DCR1 | | DCR2 | |
|---|---|---|---|---|
| | Cells undergone apoptosis | Type 1 cell | Cells undergone apoptosis | Type 1 cell |
| Healthy cells | 13 | 3 (23%) | 12 | 6 (50%) |
| Cancer cells with high Apaf | 43 | 2 (5%) | 51 | 3 (6%) |
| Cancer cells with high DR | 55 | 33 (60%) | 48 | 30 (63%) |

(ii) Selective killing of cancer cells under low affinity death ligand induction

We next considered the case when TRAIL affinity for death receptors DR4/DR5 is low ($K_d \sim 10$ μM). The affinity of the death ligand for decoy receptors was assumed to be similar to that for death receptors. Such low affinity death ligands might remain protective even for normal cells that have significant expression levels of both death and decoy receptors (as well as other membrane-proximal apoptotic molecules). In Figure 9, apoptotic activation data is shown for healthy and cancer cells. Irrespective of the pathway (type 1/type 2) of activation, we observed selective killing of cancer cells while healthy cells were mostly protected. For DL = 5, the average time-to-death and cell-to-cell variability was estimated to be approximately two to three fold higher in healthy cells than in cancer cells.

In Figure 9, we present quantitative estimations of type 1 and type 2 activations (based on average over 64 single cells). Consistent with the data shown in Figure 9, for certain types of cancer (or for some cells within a tumor population) a mixed type 1/type 2 activation phenotype will be the mode of activation. Our type 2 activation data (for both high and low affinity ligands) should be relevant for apoptosis induction in certain neural cancers as recent experimental data indicates selective activation of intrinsic apoptosis in glioblastoma cells by a combined treatment of TRAIL and embelin [41]. Strong caspase 9 activation in those experiments seems to indicate activation through the type 2 pathway. From our simulations for low affinity ligands, single cell data for caspase 9 and caspase 3 activation (Figure 10) provide further insights into selective killing of cancer cells. In cancer cells cell-to-cell variability in caspase 3 activation is markedly reduced compared with that in healthy cells.

A quantitative scoring approach can be developed for choosing optimal death ligand concentration needed for selective killing of cancer cells. In this scoring approach quantitative assessment can be done for various key factors such as: (i) number of cells undergoing apoptotic activation in a given simulation time (highest score of 10 for 100% apoptosis) (ii) average time-to-death (highest score of 5 for $T_d = 0$ to lowest 0 when $T_d$ equals the maximum simulation time), (iii) cell-to-cell variability in $T_d$ (the case with cell-to-cell variability equals to half-maximal simulation time is assigned a score of 0 and no cell-to-cell variability is assigned a maximum score of 5), (iv) number of apoptosis in corresponding healthy cells (highest score of 10 for 0% apoptosis). When we applied this quantitative method for cancer cells equipped with high death receptor (DR = 100), DL = 5-10 turned out to be the optimal DL level. The following death ligand levels DL = 2, 5 and 10 were analyzed based on data shown in Figure 7. Such a quantitative data analysis approach can be more generally applicable, such as to find optimal ligand avidity or optimal combinatorial treatment options, for selective apoptotic activation in cancer cells.

**Table 2.** Quantitative scoring for choosing optimal death ligand concentration

|      | DL = 2 | DL = 5 | DL = 10 |
|------|--------|--------|---------|
| DCR1 | 17.9   | 20     | 20      |
| DCR2 | 18.7   | 21.4   | 20.6    |

We also studied death ligands having increased affinity for decoy receptors than death receptors to probe whether such ligands could selectively induce apoptosis in cancer cells. When expression of death receptors (DR4 / DR5) is significant in healthy cells (cells that will be under the action of chemotherapeutic drugs), but still the DR/ DCR ratio is larger in cancer cells compared to that in healthy cells, engineered ligands that target the death receptors with lower than usual affinity (∼ nanomolar) might allow minimal cytotoxicity in normal cells. The affinity for decoy receptors remains in the usual nanomolar range so that most ligands bind to decoy receptors on normal cells instead of death receptors. Even TRAIL (the natural death ligand) might bind with DCR2 with slightly higher effective affinity (than that for DR4/DR5) due to elongated and rigid extracellular structure of DCR2 [28]. In our simulations, for healthy cells we assumed DR = 10 and DCR = 20, whereas, for cancer cells DCR is still 20 molecules but DR / DCR ≥ 1. When the type 2 pathway is susceptible for activation, DR ∼ 20 (DR / DCR ∼ 1) and DL ∼ 10 are sufficient for significant apoptotic induction in cancer cells (Figure 11). In a similar manner, when high levels of anti-apoptotic proteins (such as Bcl2) makes the type 2 pathway particularly resistant but DR4 / DR5 expression is high (either constitutively or induced by genetic or other mechanisms), then the type 1 activation can be achieved in a selective manner as shown in Figure 11. However, such a strategy, based on lower affinity ligands for the death receptor, would fail if the decoy receptor expression on cancer cells is higher than that of the death receptor.

**Discussion**

In many cancer cells, over-expression of pro- and anti-apoptotic molecules may make them particularly prone to either type 1 or type 2 activation. For example, high over-expression level of XIAP makes the type 2 pathway particularly resistant. In contrast, low expression level of death receptors and high cFLIP level may preferentially activate the type 2 pathway in a cancer cell. However, increased concentration of DL increases the type 1 activation in most cases. Increase in type 1 activation results from increased caspase 8 activation that compensates for the lower binding affinity of caspase 8- procaspase 3 than that of caspase 8-Bid. For some types of cancer, increased DL concentration will presumably allow a type 2 to type 1 transition (as also observed in our earlier studies for healthy cells [15]). In our simulations, the type 2 activation in cancer cells exhibits its characteristic all-or-none type behavior with large cell-to-cell variability [17, 33]. Therefore, type 2 to type 1 transition seems to allow a stochastic-to-deterministic change in apoptotic activation. In certain cancers, such a stochastic-to-deterministic change (through type 2 to type 1 transition) will be possible to achieve by combinatorial treatments, such as by the combined action of TRAIL and XIAP inhibitors [3, 14, 18]. Incorporation of decoy receptors in our current study introduces additional complexity into the type 1/ type 2 choice problem in apoptotic activation. Significant amount of decoy receptor expression on healthy cells, as observed in some early studies [22, 23], largely limited the apoptotic activation (often with a type 2 phenotype) under low level of death ligand induction. In a similar manner, decoy receptors on cancer cells significantly diminished apoptotic activation (especially when the death ligand concentration was low) [31]. The DR / DCR ratio emerged as key parameter controlling apoptotic activation and cell-to-cell variability. Our Monte Carlo simulations could provide a clear insight into the mechanisms of two different types of decoy receptors. For the case of DCR1, the decoy receptors are pre-clustered (possibly in lipid raft domains [2]), and, recruits death ligands in those clusters. Thus, upon death ligand induction death receptors cluster separately but the death receptors now compete (for ligands) with already clustered decoy receptors. For DCR2, death ligand induction leads to hetero-clusters of death (only DR5) and decoy receptors resulting in reduced clustering / proximity of procaspase 8 molecules (as only the death receptors but not the decoy receptors recruit procaspse 8). Under conditions of low level of available death adaptor proteins anti-apoptotic effect of DCR2 is possibly stronger than that of DCR1. Stronger anti-apoptotic effect of DCR1 might also be observed in some other situations. Hence, it might be possible to modulate the anti-apoptotic effect of decoy receptors through variation in the DCR1/DCR2 ratio on a given cell type.

How to selectively target the apoptotic pathways in cancer cells, that would spare normal cells, has been a key question in cancer biology and cancer therapy. Here, results obtained from our computational studies elucidate mechanisms of selective killing of cancer cells under TRAIL therapy. We show that inherent differences in the DR/DCR ratio (death receptor to decoy receptor ratio) between cancer and healthy cells may provide a mechanism for TRAIL induced selective killing of cancer cells. The DR/DCR ratio modulates cell-to-cell variability in apoptotic activation generating crucial differences in apoptotic activation between cancer and healthy cells. It might also be possible to activate other death ligand mediated pathways for certain cancer types (such as Fas-FasL for brain tumors [40]). For the case of Fas-FasL, the relative abundance of death receptors on tumor and healthy cells presumably provides a mechanism for selectivity [40, 42]. In some situations, the death receptor (and/or other pro-apoptotic molecules in the membrane) expression can even be induced in cancer cells by genetic or other techniques [40], leading to selective apoptotic activation in cancer cells. In a more general manner, inherent (or induced) differences in expression levels of receptor and/or other molecules in the membrane proximal module and their membrane organization [46], between cancer and healthy cells, may generate selectivity in cancer cell targeting. In addition, such inherent (or induced) differences may exist in downstream signaling modules (such as high Apaf expression in certain neural cancer cells [44]) contributing to robust selectivity in death ligand induced cancer cell apoptosis. In the cases where high level of anti-apoptotic proteins are also present (such as cFLIP, Bcl2 and XIAP), combination therapy with TRAIL and inhibitor molecules will help achieve high selectivity in cancer cell apoptosis [3, 18]. One may also consider affinity variant recombinant death ligands in the following manners: (i) low affinity ligands for specific death and decoy receptors, (ii) ligands selective for specific decoy receptors (low affinity for death receptors) and (iii) ligands selective for specific death receptors (low affinity for decoy receptors). In this study, we considered the possibility of using low affinity recombinant TRAIL as a strategy to enhance selectivity in cancer cell killing. Based on our results, it is expected that such low affinity ligands will allow binding of majority of death ligands to decoy receptors on healthy cells leading to fewer apoptosis of those cells. In this context, large cell-to-cell variability (in apoptotic activation) and slow activation can be protective for healthy cells. Therefore, cell-to-cell variability provides new insights into the problem of understanding the mechanism of selective killing of cancer cells.

Selective targeting of cancer cells might be possible in various possible manners. In cancer cells equipped with high level of death receptors (TRAIL receptors DR4/DR5) and low level decoy receptors activating the type 1 apoptotic pathway might be suitable for selective killing. In some other cancer cells, with slightly defective membrane proximal signaling module (such as due to low level of death adaptor proteins) selective apoptotic activation through the type 2 pathway might be possible by death ligand induction. If Bcl2 is also overexpressed in such cancer cells then selective activation can be achieved under combined action of TRAIL and Bcl2 inhibitor. In yet another scenario, if cancer cells equipped with high concentrations of anti-apoptotic Bcl2 proteins also have strongly over-expressed Bid and Bax (pro-apoptotic) molecules, then removal of Bcl2 inhibition by inhibitory small molecules could be sufficient for selective activation of the cancer cell type 2 pathway. *In silico* approaches may help develop the optimal strategy for targeting the apoptotic pathway in specific cancer cells. Information regarding clonal heterogeneity in a tumor population [35, 36] and single cell genomic/proteomic data [47] can be utilized for designing Monte Carlo based *in silico* studies of apoptotic activation. Single cell activation data, as obtained from our Monte Carlo simulations, can be analyzed utilizing a quantitative scoring method in order to compare potential options for apoptosis induction in cancer cells. In such a scoring approach quantitative assessment can be based on several key factors such as: maximize cancer cells killing while keeping healthy cell killing at a minimal level, minimize cell-to-cell variability and time-to-death in cancer cell killing. In this study, we have applied a scoring method for low affinity death ligand induction (for apoptosis activation data in Figure 7) to find the optimal death ligand level needed for selectively inducing cancer cell apoptosis. Such a quantitative data analysis method can also be useful for analyzing single cell activation data in experimental and clinical settings.

**References**

1.  Falschlehner, C., et al., *TRAIL signalling: decisions between life and death.* Int J Biochem Cell Biol, 2007. **39**(7-8): p. 1462-75.
2.  Shirley, S., A. Morizot, and O. Micheau, *Regulating TRAIL receptor-induced cell death at the membrane : a deadly discussion.* Recent Pat Anticancer Drug Discov, 2011. **6**(3): p. 311-23.
3.  Kurita, S., et al., *Hedgehog inhibition promotes a switch from Type II to Type I cell death receptor signaling in cancer cells.* PLoS One, 2011. **6**(3): p. e18330.
4.  Picarda, G., et al., *TRAIL receptor signaling and therapeutic option in bone tumors: the trap of the bone microenvironment.* Am J Cancer Res, 2012. **2**(1): p. 45-64.
5.  Merino, D., et al., *Differential inhibition of TRAIL-mediated DR5-DISC formation by decoy receptors 1 and 2.* Mol Cell Biol, 2006. **26**(19): p. 7046-55.
6.  Rathmell, J.C. and C.B. Thompson, *Pathways of apoptosis in lymphocyte development, homeostasis, and disease.* Cell, 2002. **109 Suppl**: p. S97-107.
7.  Elmore, S., *Apoptosis: a review of programmed cell death.* Toxicol Pathol, 2007. **35**(4): p. 495-516.
8.  Takeda, K., et al., *Involvement of tumor necrosis factor-related apoptosis-inducing ligand in surveillance of tumor metastasis by liver natural killer cells.* Nat Med, 2001. **7**(1): p. 94-100.
9.  Quintana, E., et al., *Efficient tumour formation by single human melanoma cells.* Nature, 2008. **456**(7222): p. 593-8.
10. Smyth, M.J., et al., *Nature's TRAIL--on a path to cancer immunotherapy.* Immunity, 2003. **18**(1): p. 1-6.
11. Di Carlo, M., *Beta amyloid peptide: from different aggregation forms to the activation of different biochemical pathways.* Eur Biophys J, 2010. **39**(6): p. 877-88.
12. Picone, P., et al., *Abeta oligomers and fibrillar aggregates induce different apoptotic pathways in LAN5 neuroblastoma cell cultures.* Biophys J, 2009. **96**(10): p. 4200-11.
13. Fossati, S., J. Ghiso, and A. Rostagno, *TRAIL death receptors DR4 and DR5 mediate cerebral microvascular endothelial cell apoptosis induced by oligomeric Alzheimer's Abeta.* Cell Death Dis, 2012. **3**: p. e321.
14. Raychaudhuri, S. and S.C. Raychaudhuri, *Monte Carlo Study Elucidates the Type 1/Type 2 Choice in Apoptotic Death Signaling in Healthy and Cancer Cells.* Cells, 2013. **2**(2): p. 361-392.
15. Raychaudhuri, S. and S.C. Raychaudhuri, *Death ligand concentration and the membrane proximal signaling module regulate the type 1/type 2 choice in apoptotic death signaling.* Syst Synth Biol, 2014. **8**(1): p. 83-97.
16. Certo, M., et al., *Mitochondria primed by death signals determine cellular addiction to antiapoptotic BCL-2 family members.* Cancer Cell, 2006. **9**(5): p. 351-65.
17. Skommer, J., T. Brittain, and S. Raychaudhuri, *Bcl-2 inhibits apoptosis by increasing the time-to-death and intrinsic cell-to-cell variations in the mitochondrial pathway of cell death.* Apoptosis, 2010. **15**(10): p. 1223-33.

18. Skommer, J., et al., *Nonlinear regulation of commitment to apoptosis by simultaneous inhibition of Bcl-2 and XIAP in leukemia and lymphoma cells.* Apoptosis, 2011. **16**(6): p. 619-26.
19. Raychaudhuri, S. and S.C. Das, *Low probability activation of Bax / Bak can induce selective killing of cancer cells by generating heterogeneoity in apoptosis.* Journal of Healthcare Engineering, 2013. **4**: p. 47-66.
20. Breen, L., et al., *Investigation of the role of p53 in chemotherapy resistance of lung cancer cell lines.* Anticancer Res, 2007. **27**(3A): p. 1361-4.
21. Lin, T., et al., *Long-term tumor-free survival from treatment with the GFP-TRAIL fusion gene expressed from the hTERT promoter in breast cancer cells.* Oncogene, 2002. **21**(52): p. 8020-8.
22. Pan, G., et al., *An antagonist decoy receptor and a death domain-containing receptor for TRAIL.* Science, 1997. **277**(5327): p. 815-8.
23. Sheridan, J.P., et al., *Control of TRAIL-induced apoptosis by a family of signaling and decoy receptors.* Science, 1997. **277**(5327): p. 818-21.
24. *Apoptosis in Health and Diseases*. First Paperback ed. 2010, Cambridge UK: Cambridge University Press.
25. Ngamkitidechakul, C., et al., *Antitumour effects of Phyllanthus emblica L.: induction of cancer cell apoptosis and inhibition of in vivo tumour promotion and in vitro invasion of human cancer cells.* Phytother Res, 2010. **24**(9): p. 1405-13.
26. Guicciardi, M.E. and G.J. Gores, *Life and death by death receptors.* FASEB J, 2009. **23**(6): p. 1625-37.
27. Hua, F., et al., *Effects of Bcl-2 levels on Fas signaling-induced caspase-3 activation: molecular genetic tests of computational model predictions.* J Immunol, 2005. **175**(2): p. 985-95.
28. Schneider, P., et al., *Characterization of two receptors for TRAIL.* FEBS Lett, 1997. **416**(3): p. 329-34.
29. Meng, X.W., et al., *High cell surface death receptor expression determines type I versus type II signaling.* J Biol Chem, 2011. **286**(41): p. 35823-33.
30. Gu, C., et al., *A trigger model of apoptosis induced by tumor necrosis factor signaling.* BMC Syst Biol, 2011. **5 Suppl 1**: p. S13.
31. Sanlioglu, A.D., et al., *Surface TRAIL decoy receptor-4 expression is correlated with TRAIL resistance in MCF7 breast cancer cells.* BMC Cancer, 2005. **5**: p. 54.
32. Fricker, N., et al., *Model-based dissection of CD95 signaling dynamics reveals both a pro- and antiapoptotic role of c-FLIPL.* J Cell Biol, 2010. **190**(3): p. 377-89.
33. Raychaudhuri, S., et al., *Monte Carlo simulation of cell death signaling predicts large cell-to-cell stochastic fluctuations through the type 2 pathway of apoptosis.* Biophys J, 2008. **95**(8): p. 3559-62.
34. Bagci, E.Z., et al., *Bistability in apoptosis: roles of bax, bcl-2, and mitochondrial permeability transition pores.* Biophys J, 2006. **90**(5): p. 1546-59.
35. Notta, F., et al., *Evolution of human BCR-ABL1 lymphoblastic leukaemia-initiating cells.* Nature, 2011. **469**(7330): p. 362-7.
36. Anderson, K., et al., *Genetic variegation of clonal architecture and propagating cells in leukaemia.* Nature, 2011. **469**(7330): p. 356-61.

37. Hassan, M., et al., *Apoptosis and molecular targeting therapy in cancer.* Biomed Res Int, 2014. **2014**: p. 150845.
38. Sun, X.M., et al., *Bcl-2 and Bcl-xL inhibit CD95-mediated apoptosis by preventing mitochondrial release of Smac/DIABLO and subsequent inactivation of X-linked inhibitor-of-apoptosis protein.* J Biol Chem, 2002. **277**(13): p. 11345-51.
39. Maas, C., et al., *Smac/DIABLO release from mitochondria and XIAP inhibition are essential to limit clonogenicity of Type I tumor cells after TRAIL receptor stimulation.* Cell Death Differ, 2010. **17**(10): p. 1613-23.
40. Ho, I.A., W.H. Ng, and P.Y. Lam, *FasL and FADD delivery by a glioma-specific and cell cycle-dependent HSV-1 amplicon virus enhanced apoptosis in primary human brain tumors.* Mol Cancer, 2010. **9**: p. 270.
41. Siegelin, M.D., T. Gaiser, and Y. Siegelin, *The XIAP inhibitor Embelin enhances TRAIL-mediated apoptosis in malignant glioma cells by down-regulation of the short isoform of FLIP.* Neurochem Int, 2009. **55**(6): p. 423-30.
42. Liu, X., et al., *Death receptor regulation and celecoxib-induced apoptosis in human lung cancer cells.* J Natl Cancer Inst, 2004. **96**(23): p. 1769-80.
43. Yoshida, T., et al., *Kaempferol sensitizes colon cancer cells to TRAIL-induced apoptosis.* Biochem Biophys Res Commun, 2008. **375**(1): p. 129-33.
44. Johnson, C.E., et al., *Differential Apaf-1 levels allow cytochrome c to induce apoptosis in brain tumors but not in normal neural tissues.* Proc Natl Acad Sci U S A, 2007. **104**(52): p. 20820-5.
45. Hu, R., et al., *Embelin induces apoptosis through down-regulation of XIAP in human leukemia cells.* Med Oncol, 2011. **28**(4): p. 1584-8.
46. Marconi, M., et al., *Constitutive localization of DR4 in lipid rafts is mandatory for TRAIL-induced apoptosis in B-cell hematologic malignancies.* Cell Death Dis, 2013. **4**: p. e863.
47. Yang, Y., et al., *Parallel single cancer cell whole genome amplification using button-valve assisted mixing in nanoliter chambers.* PLoS One, 2014. **9**(9): p. e107958.

**Figures**

Figure 1. A schematic for the apoptotic death signaling network showing membrane proximal signaling module and type 1/type 2 pathways.

Figure 2. Death ligand induced apoptotic activation (a) and type 1/type 2 choice (b) for cancer cells equipped with over-expressed Bcl2 and XIAP. In our simulations, we used: Bcl2 = 750 nM (10 fold overexpression), XIAP = 90 nM (3 fold overexpression), Bid = 125 nM (5 fold overexpression), Bax = 415 nM (5 fold overexpression). Overexpression is defined with respect to concentrations used in simulating apoptosis in healthy cells. Death ligand was varied in the following manner: DL = 2 and 10 (for a 1.2 x 1.2 $\mu m^2$ cell surface area); DR = 10 and DCR = 20. Data is shown for two different scenarios of decoy receptor engagement (DCR1 and DCR2). Increased death ligand concentration resulted in increased apoptotic activation and an increase in type 1 activation fraction.

Figure 3. Death ligand induced apoptotic activation (a) and type 1/type 2 choice (b) for cancer cells equipped with over-expressed Bcl2 and cFLIP but decreased death adaptor proteins. In our simulations, we used: Bcl2 = 750 nM (10 fold overexpression), cFLIP = 60 nM (2 fold overexpression), adaptor protein = 10 nM (3 fold lower expression), Bid = 125 nM (5 fold overexpression), Bax = 415 nM (5 fold overexpression). Death ligand was varied in the following manner: DL = 2, 10 and 100 (for a 1.2 x 1.2 $\mu m^2$ area); DR = 10 and DCR = 20. Data is shown for two different scenarios of decoy receptor engagement (DCR1 and DCR2). Increased death ligand concentration resulted in increased apoptotic activation and an increase in type 1 activation fraction.

Figure 4. Single cell caspase 3 activation (under death ligand induction) in cancer cells equipped with over-expressed Bcl2 and cFLIP but decreased death adaptor proteins. In our simulations, we used: Bcl2 = 750 nM (10 fold overexpression), cFLIP = 60 nM (2 fold overexpression), adaptor protein = 10 nM (3 fold lower expression), Bid = 125 nM (5 fold overexpression), Bax = 415 nM (5 fold overexpression). Death ligand was varied in the following manner: DL = 2, 10 and 100 (for a 1.2 x 1.2 $\mu m^2$ area); DR = 10 and DCR = 20. Data is shown for two different scenarios of decoy receptor engagement (DCR1 and DCR2). Each curve corresponds to apoptosis activation for a single cell (Monte Carlo run). Data is shown for 16 representative single cells. Large cell-to-cell variability and all-or-none behavior is observed indicating type 2 activation.

Figure 5. Death ligand induced apoptotic activation (a) and type 1/type 2 choice (b) for healthy cells. Healthy cells were assumed to have low level of adaptor proteins (= 10 nM). Death ligand was varied in the following manner: DL = 2, 10 and 100 (for a 1.2 x 1.2 $\mu m^2$ area); DR = 20 and DCR = 20. Data is shown for two different scenarios of decoy receptor engagement. Increased death ligand concentration resulted in increased apoptotic activation and a type 2 to type 1 transition.

Figure 6. Clustering of death (shown in red) and decoy receptors (shown in green) upon death ligand engagement. Data is shown for two different scenarios of decoy receptor engagement: DCR1 (a) and DCR2 (b). Clustering on a representative single cell (Monte Carlo run) is shown for 4 different time-points (t = $10^6$, $2 \times 10^6$, $5 \times 10^6$ $10^7$). DL = 100, DR = 100 and DCR = 100 was used in our simulations.

Figure 7. Death ligand induced apoptotic activation (a) and type 1/type 2 choice (b) for cancer cells equipped with over-expressed Bcl2 family anti- and pro-apoptotic proteins. In our simulations, we used: Bcl2 = 750 nM (10 fold overexpression), Bid = 125 nM (5 fold overexpression), Bax = 415 nM (5 fold overexpression). Death ligand was varied in the following manner: DL = 2, 10 and 100 (for a 1.2 x 1.2 $\mu m^2$ area); DR = 10 and DCR = 20. Data is shown for two different scenarios of decoy receptor engagement. Increased death ligand concentration resulted in increased apoptotic activation and a type 2 to type 1 transition.

Figure 8. Death ligand induced selective activation of the apoptotic pathway in cancer cells. Cancer cells were assumed to have higher level of death adaptor proteins (30 molecules compared with 10 in healthy cells). For cancer cells susceptible to type 2 activation, high expression level of Apaf (100 nM) was used in our simulations. Slightly higher expression level of death receptors (20 molecules compared with 10 in healthy cells) was also used. For cancer cells susceptible to type 1 activation, high expression level of death receptors (100 molecules) was used, but the expression level of Bcl2 was also assumed to be high (750 nM). In our simulations we also used Bid = 125 nM (5 fold overexpression) and Bax = 415 nM (5 fold overexpression) for all types of cancer cells. Death ligand induced apoptotic activation is shown to be significantly higher in cancer cells (DL = 2 molecules). Data is shown for two different scenarios of decoy receptor engagement: (a) DCR1 and (b) DCR2.

Figure 9. Selective apoptotic activation in healthy and cancer cells under induction of low affinity ligands for the death and decoy receptors ($K_a = 10^5$ $M^{-1}$). In the membrane module of cancer cells, we assumed higher level of adaptor proteins (30 molecules in cancer cells and 10 in healthy cells) but lower level of cFLIP (10 molecules in cancer cells compared 30 in healthy cells). For cancer cells susceptible to type 2 activation, high expression level of Apaf (100 nM) was used in our simulations. Slightly higher expression level of death receptors (20 molecules compared with 10 in healthy cells) was also used. For cancer cells susceptible to type 1 activation, high expression level of death receptors (100 molecules) was used, but the expression level of Bcl2 was also assumed to be high (750 nM). In our simulations we also used Bid = 125 nM (5 fold overexpression) and Bax = 415 nM (5 fold overexpression) for all types of cancer cells. Death ligand induced apoptotic activation is shown to be significantly higher in cancer cells (DL = 2, 5 and 10 molecules were used in simulations). Data is shown for two different scenarios of decoy receptor engagement: (a) DCR1 and (b) DCR2.

Figure 10. Single cell caspase 3 (upper panel) and caspase 9 (lower panel) activation in healthy and cancer cells under induction of low affinity ligands for the death and decoy receptors ($K_a = 10^5$ $M^{-1}$). Data is shown for DL = 5. Other parameters are same as in Figure 9. Each curve corresponds to apoptosis activation for a single cell (Monte Carlo run). Data is shown for 8 representative single cells for two different scenarios of decoy receptor engagement: (a) DCR1 and (b) DCR2.

Figure 11. Selective apoptotic activation in cancer cells under induction of low affinity ligands for the death receptors ($K_a = 10^7$ $M^{-1}$). In the membrane module of cancer cells, we assumed higher level of adaptor proteins (30 molecules in cancer cells and 10 in healthy cells) but lower level of cFLIP (10 molecules in cancer cells compared 30 in healthy cells). For cancer cells susceptible to type 2 activation, high expression level of Apaf (100 nM) was used in our simulations. Slightly higher expression level of death receptors (20 molecules compared with 10 in healthy cells) was also used. For cancer cells susceptible to type 1 activation, high expression level of death receptors (100 molecules) was used, but the expression level of Bcl2 was also assumed to be high (750 nM). In our simulations we also used Bid = 125 nM (5 fold overexpression) and Bax = 415 nM (5 fold overexpression) for all types of cancer cells. Death ligand induced apoptotic activation is shown to be significantly higher in cancer cells (DL = 10 molecules). Data is shown for two different scenarios of decoy receptor engagement: (a) DCR1 and (b) DCR2.

Figure 1

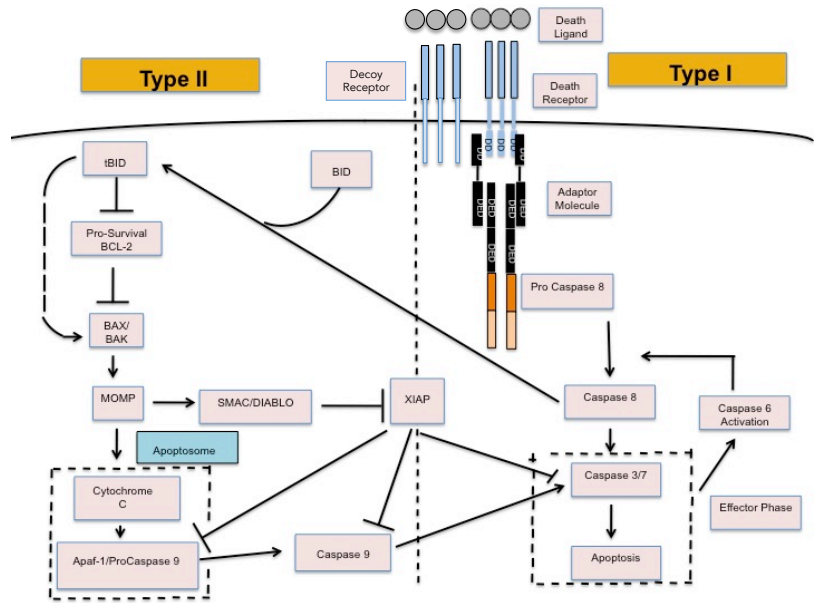

Figure 2

Figure 2a

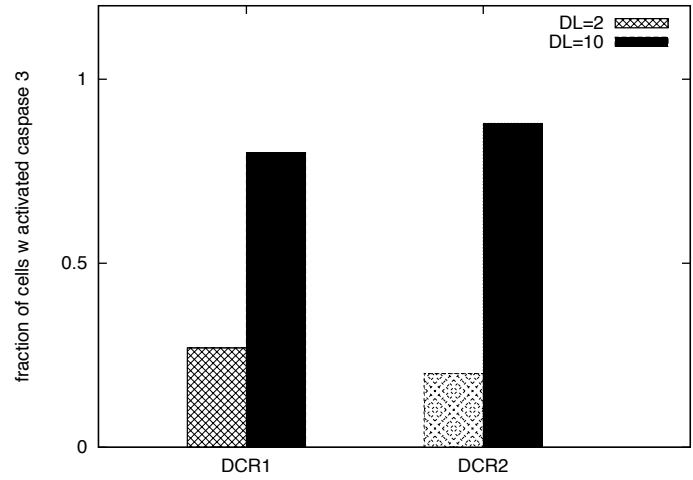

Figure 2b

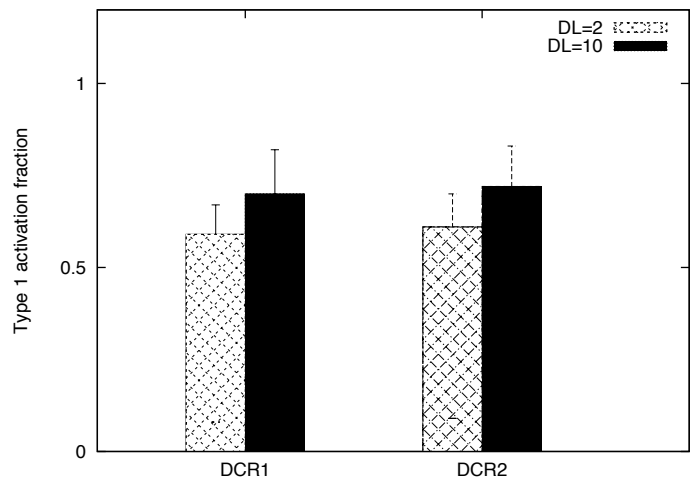

Figure 3

Figure 3a

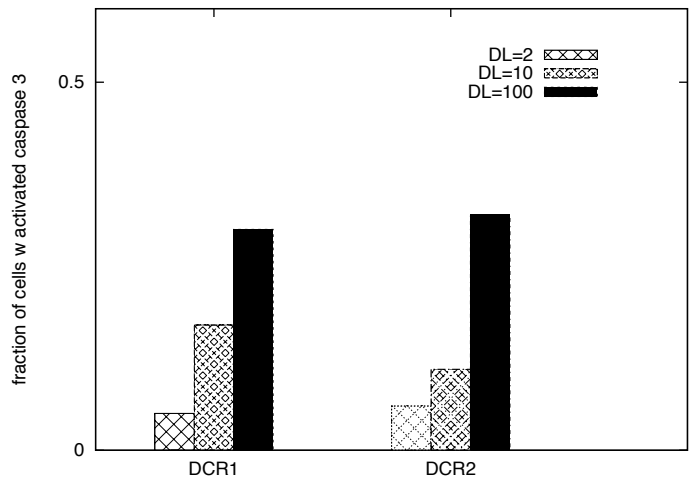

Figure 3b

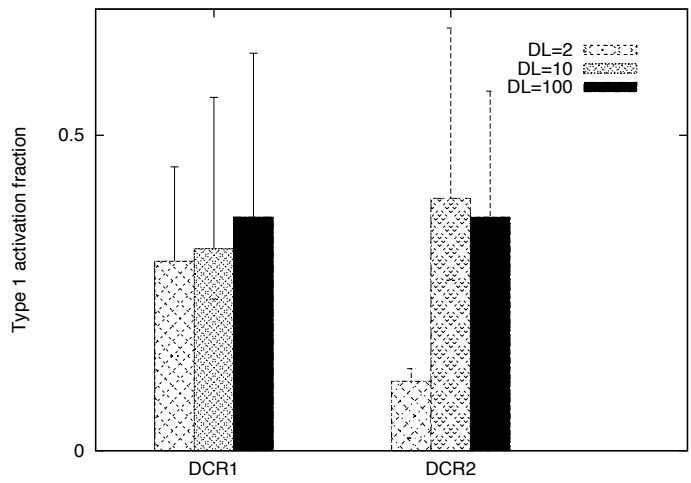

Figure 4

Figure 4a

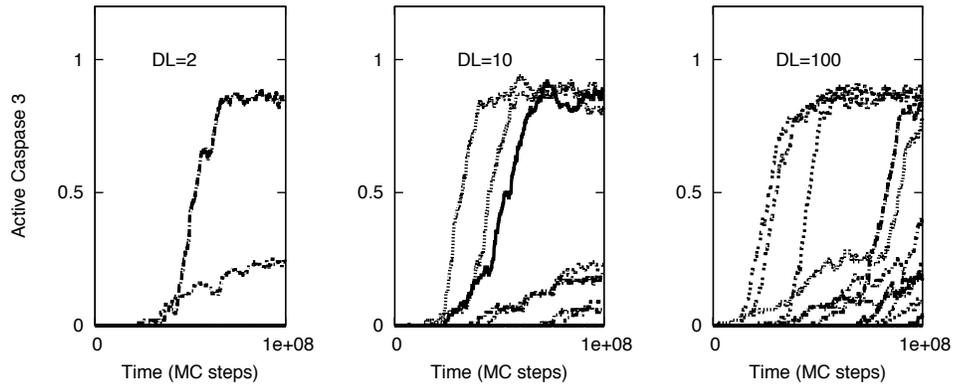

Figure 4b

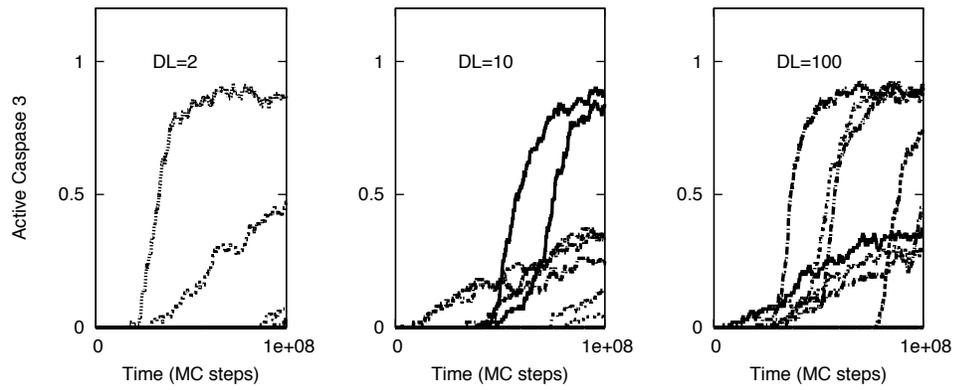

Figure 5

Figure 5a

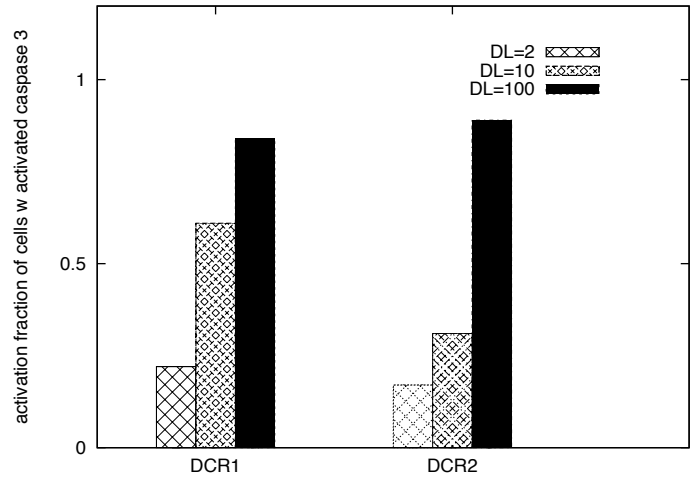

Figure 5b

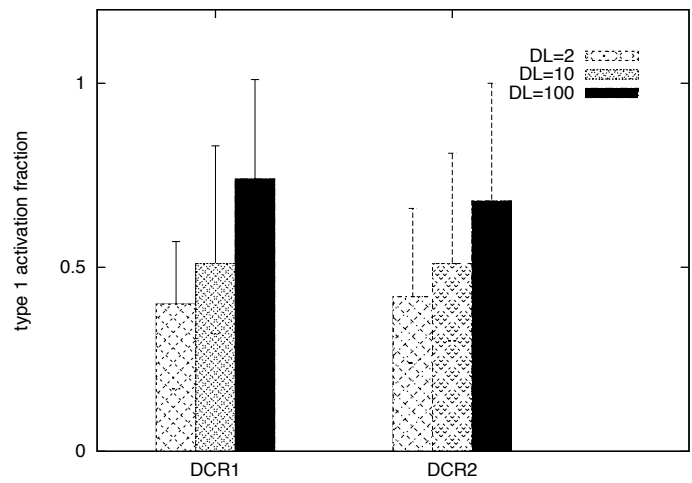

Figure 6

Figure 6a

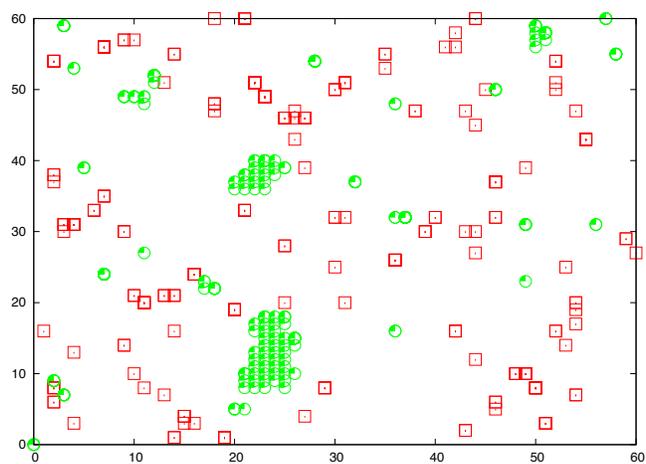

T = $10^6$ MC steps

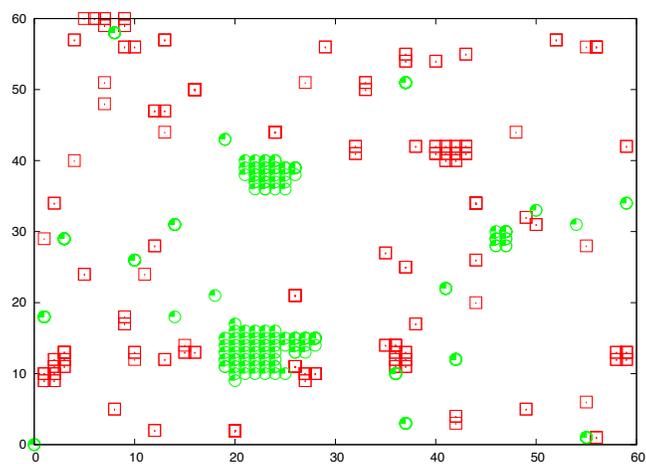

T = $2 \times 10^6$ MC steps

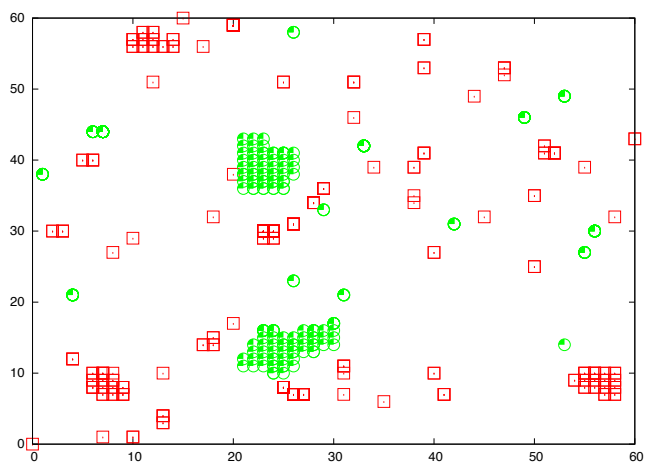

T = $5 \times 10^6$ MC steps

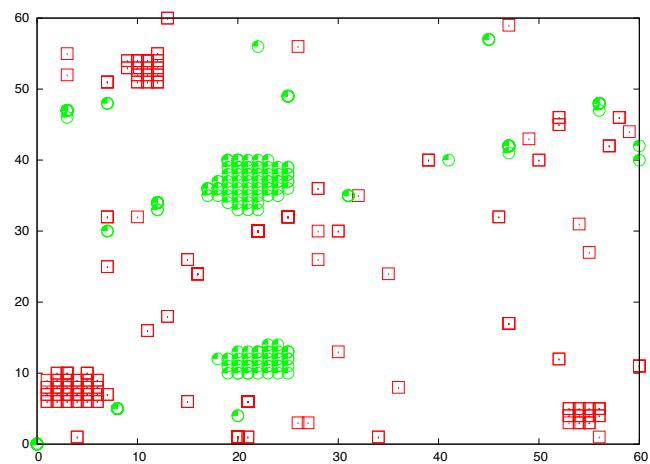

T = $10^7$ MC steps

Figure 6b

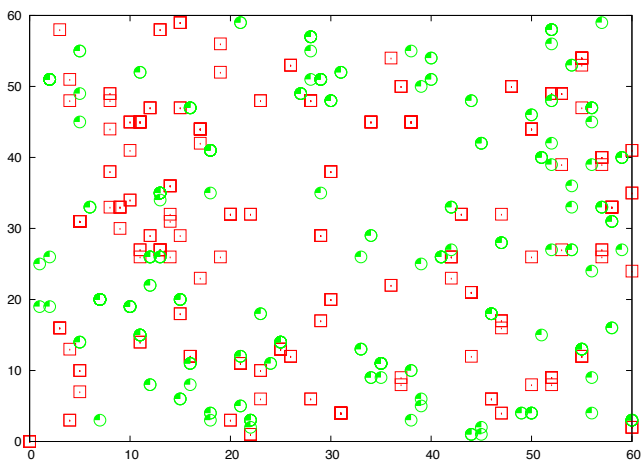

T = $10^6$ MC steps

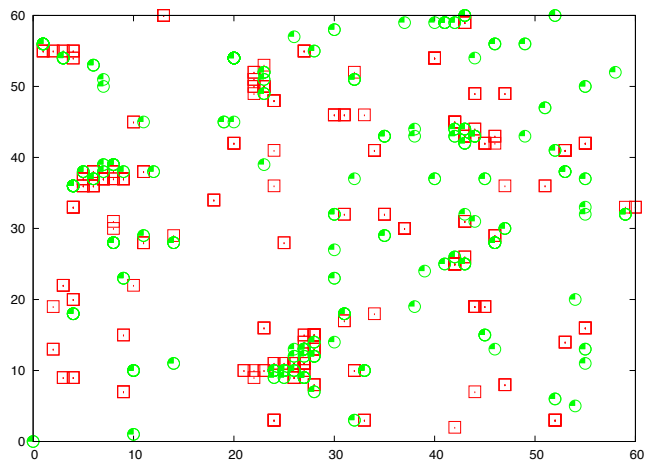

T = $2 \times 10^6$ MC steps

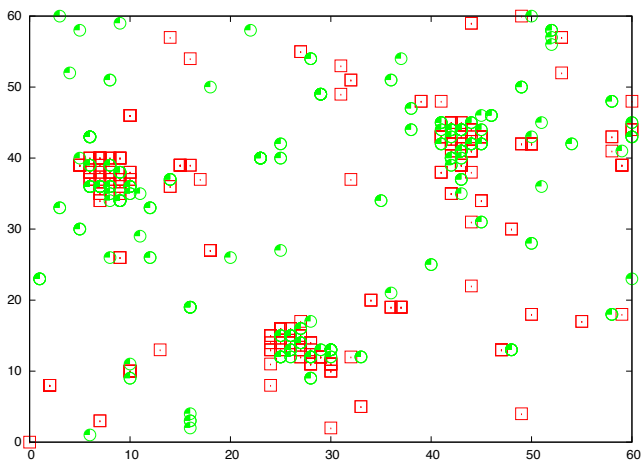

T = $5 \times 10^6$ MC steps

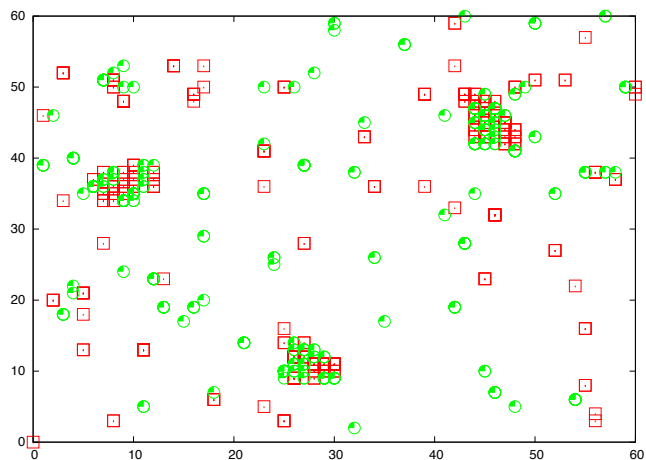

T = $10^7$ MC steps

Figure 7

Figure 7a

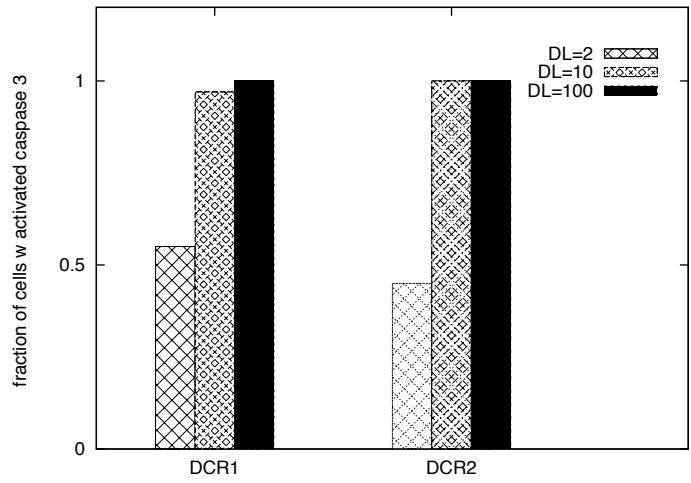

Figure 7b

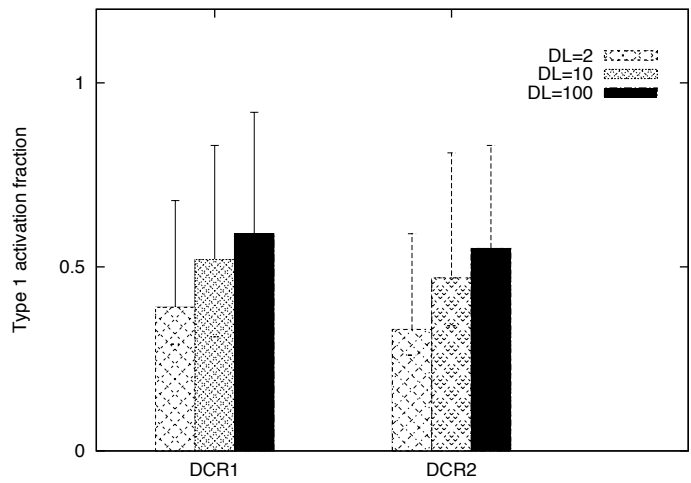

Figure 8

Figure 8a

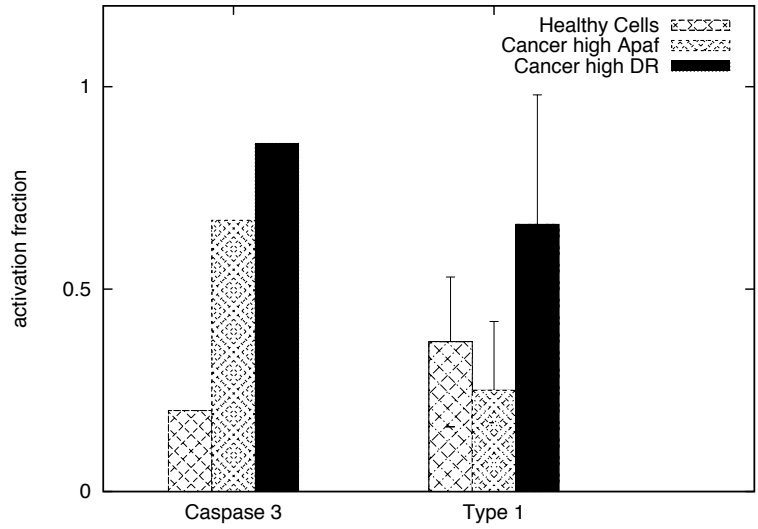

Figure 8b

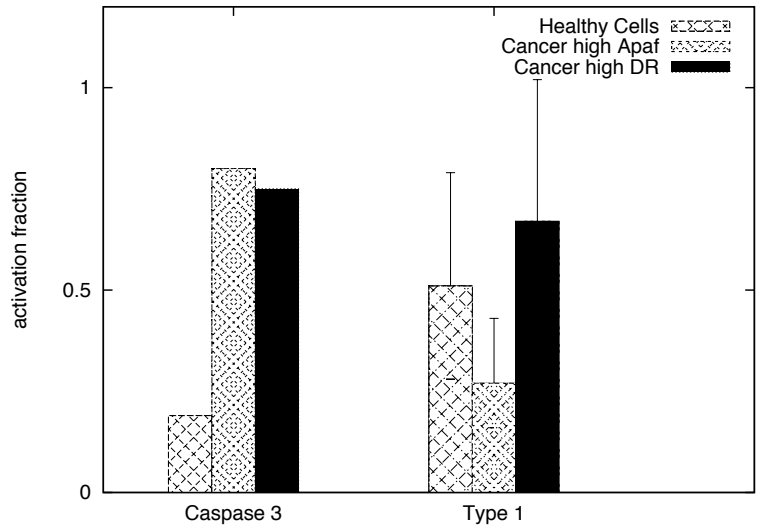

Figure 9

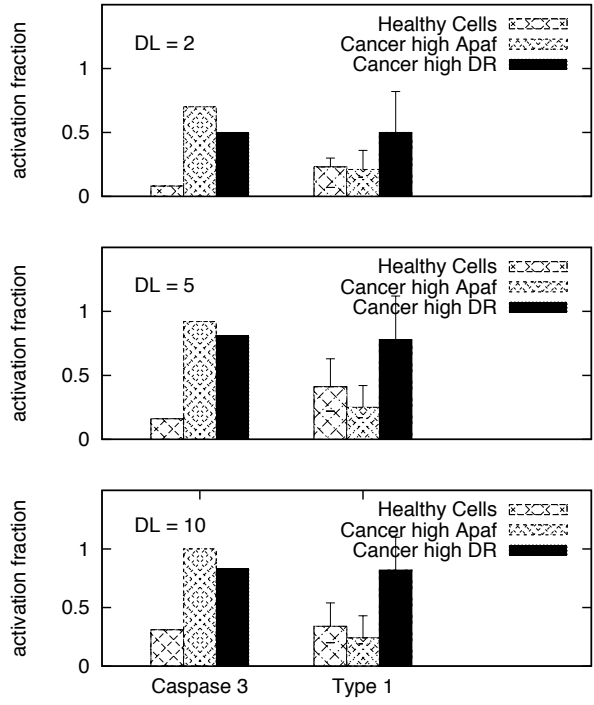

Figure 9a

Figure 9

Figure 9b

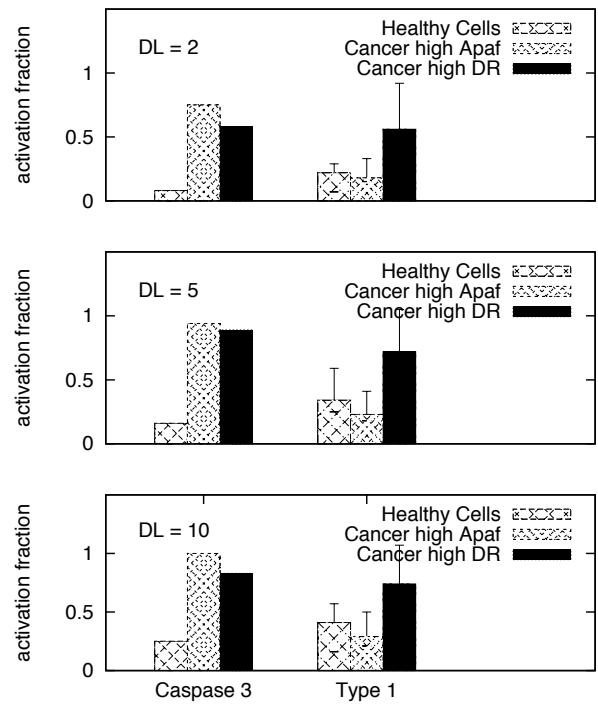

Figure 10a

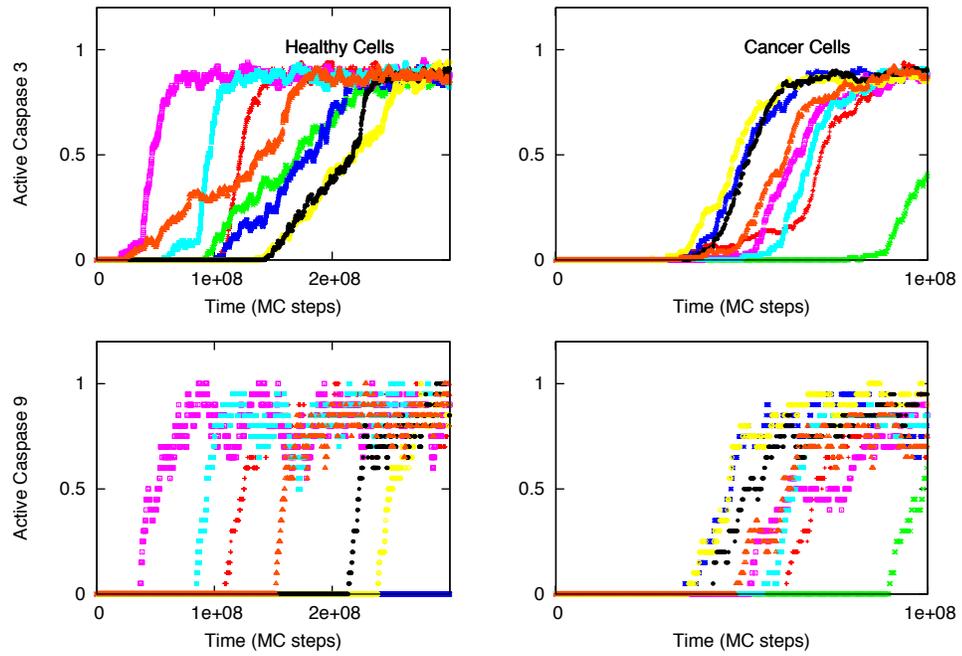

Figure 10b

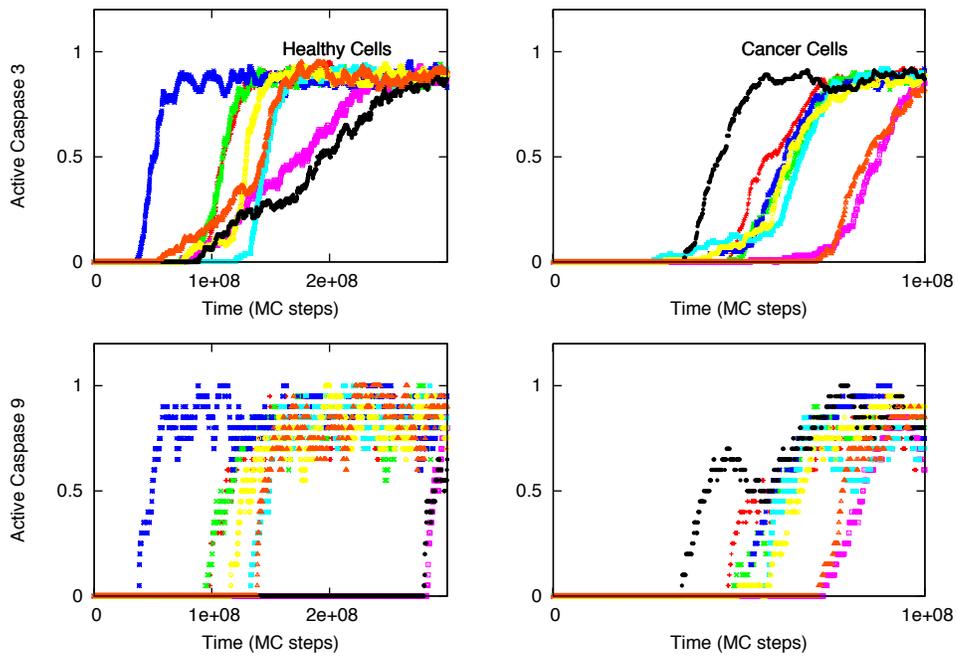

Figure 11

Figure 11a

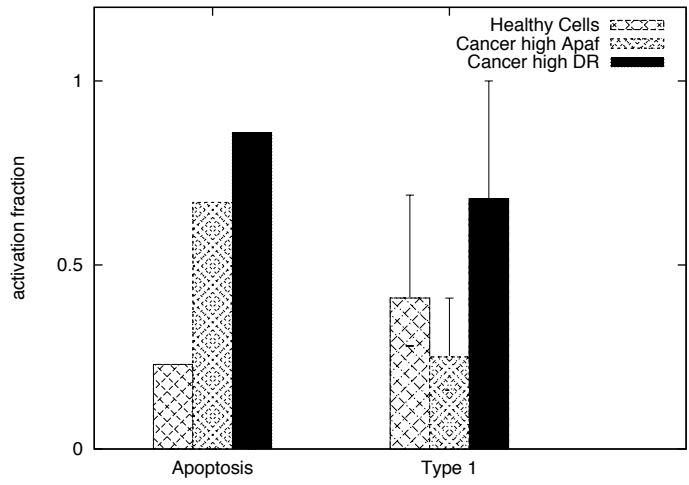

Figure 11b

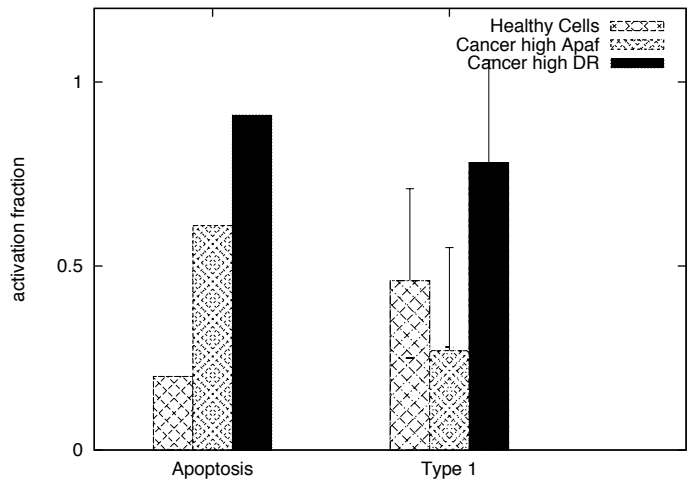